\def\year{2017}\relax
\documentclass[letterpaper]{article}
\usepackage{aaai17}
\usepackage{times}
\usepackage{helvet}
\usepackage{courier}
\begin{document}
%
\title{On Selecting a Conjunction Operation in Probabilistic Soft Logic}


\author{Vladik Kreinovich\\
Department of Computer Science\\
University of Texas at El Paso\\
vladik@cs.utep.edu
\And
Chitta Baral\\
School of Computing, Informatics and DSE\\
Arizona State University\\
chitta@asu.edu
}

\maketitle
\begin{abstract}
Probabilistic Soft Logic has been proposed and used in several applications as an efficient way to deal with inconsistency, uncertainty and relational representation. In several applications, this approach has led to an adequate description of the corresponding human reasoning. In this paper, we provide a theoretical explanation for one of the semi-heuristic choices made in this approach: namely, we explain the choice of the corresponding conjunction operations. Our explanation leads to a more general family of operations which may be used in future applications of probabilistic soft logic.
\end{abstract}

\section{Introduction and Motivation}

With the maturing of various sub-fields of AI we are now ready to combine approaches and techniques from multiple AI sub-fields to address broader issues. For example, we now have large bodies of knowledge that are available. An example such a knowledge base is ConceptNet \cite{liu2004conceptnet}.  Even in the same collection some of the knowledge may be manually curated, while parts may be automatically extracted. Sometimes all of the knowledge may be automatically obtained, such as the similarity knowledge based used in \cite{Beltagy2014} that is obtained using distributional semantics \cite{bruni2014multimodal} of natural language. There may be inconsistencies lingering inside the knowledge base. For some of the knowledge, we may be able to assign weights. Learning part of this knowledge and reasoning with such knowledge requires approaches that can handle inconsistencies, uncertainty, structured information, and most importantly the approaches need to scale.  Among the various approaches that have been proposed {\em Probabilistic Soft Logic (PSL)} \cite{Bach2010,Bach2013,Kimmig2012} stands out as it can not only handle relational structure, inconsistencies and uncertainty, thus allowing one to express rich probabilistic graphical models (such as  Hinge-loss Markov random fields), but it also seems to scale up better than its alternatives such as Markov Logic Networks \cite{Richardson2006}.  

Probabilistic soft logic (PSL) differs from most other probabilistic formalisms in that its ground atoms, instead of having binary truth values, have continuous truth values in the interval [0,1]. In the original PSL  \cite{Bach2010,Bach2013,Kimmig2012} the syntactic structure of rules and  the characterization of the logical operations have been chosen judiciously so that the space of interpretations with nonzero density forms a convex polytope. This makes inference in PSL a convex optimization problem in continuous space, which in turn allows efficient inference. The particular conjunction operation used in the above mentioned PSL is the Lukasiewicz t-norm \cite{Klir1995}. A different conjunction operation is used in \cite{Beltagy2014}, where the resulting PSL is used for semantic textual similarity.  PSL has been used in many different applications such as ones listed in\cite{Bach2010,Bach2013,Beltagy2014,Huang2012,Kimmig2012,Memory2012}. 
However, none of these works precisely justify the particular selection of conjunction operation they use, beyond listing a few implications of using those operations.

\medskip
\noindent{\bf What we plan to do.} In this paper, we provide a
theoretical explanation of the conjunction operations that are used in 
\cite{Bach2010,Bach2013,Beltagy2014,Huang2012,Kimmig2012,Memory2012} and
present a more general family of operations which may be used in future applications of 
probabilistic soft logic.

\medskip
\noindent{\bf Plan of the paper.} In this section, we recalled and contextualized
Probabilistic Soft Logic and which conjunction operations are
usually selected in this logic. In the next sections, we provide
our theoretical explanation for this selection.

\section{Rules, implications and the conjunction operation in Probabilistic Soft Logic}

\noindent{\bf The corresponding real-life problem.} In many
practical situations, we have rules $r$ of the type
$a_1,\ldots,a_n\to b$ that connect facts $a_i$ and $b$. For each
such rule $r$, we know its ``degree of importance'' $\lambda_r$:
the larger $\lambda_r$, the larger our degree of confidence in
this rule.

If we simply combine all these rules (and ignore their degrees of
importance), then usually, the resulting set of rules becomes
inconsistent.

For example, sociologists known that in elections, a person tends
to vote the same way as his friends. So, if a person B has a
friend A$_1$ who voted for an incumbent and a friend A$_2$ who
voted for a challenger, then: \begin{itemize} \item for the first
friend, the above sociological observation implies that B voted
for the incumbent, while \item for the second friend, the same
sociological observation implies that B voted against the
incumbent.
\end{itemize}

In such situations, we cannot satisfy all the rules. So, it is
reasonable to look for solutions in which an (appropriately
defined) deviation from the ideal situation -- when all the rules
are satisfied -- is the smallest possible.
\medskip

\noindent{\bf Need for probabilistic answers.} If we have two
conflicting rules, then we cannot be 100\% sure which of them is
not applicable in the current situation. If one of the rules is
more important than the other one, i.e., if its degrees of
importance $\lambda_r$ is higher ($\lambda_r>\lambda_{r'}$), then
most probably the first rule is applicable -- but it is also
possible that in this particular situation, the second rule is
applicable as well.

Thus, from the inconsistent knowledge base, we cannot extract the
exact conclusion about the corresponding facts. At best, we can
estimate the {\it probabilities} that different facts are true.
\medskip

\noindent{\bf How to deal with implication.} If $a$ implies $b$,
this means that $b$ holds in all situations in which $a$ holds --
and, maybe, in other situations as well. Thus, the probability
$p(b)$ that $b$ is true is larger than or equal to the probability
$p(a)$ that $a$ is true: $p(b)\ge p(a)$.

From this viewpoint, if we know the probabilities $p(a)$ and
$p(b)$ of two statements $a$ and $b$, and $p(a)\le p(b)$ (i.e.,
the difference $p(a)-p(b)$ is non-positive), then this inequality
is consistent with the rule $a\to b$. On the other hand, if
$p(a)>p(b)$ (i.e., if the difference $p(a)-p(b)$ is positive),
this clearly is inconsistent with the rule $a\to b$. Intuitively,
the larger the positive difference $p(a)-p(b)$, the larger the
violation. It is therefore reasonable to take $\max(p(a)-p(b),0)$
as the measure of severity of the rule's violation.

We want to minimize the overall loss of adequacy. Depending on the
rule's degree of importance $\lambda_r$, the same degree of rule's
violation may lead to different severity. In general, this
severity is a function of the rule's degree of importance
$\lambda_r$ and of its degree of violation $\max(p(a)-p(b),0)$:
$$d=s(\lambda_r,\max(p(a)-p(b),0)).$$

We want to find the probabilities for which the overall loss of
adequacy is the smallest possible. For rules $r$ of the type
$a_r\to b_r$, this means minimizing the sum
$$\sum_r s(\lambda_r,\max(p(a_r)-p(b_r),0)).\eqno{(1)}$$

In Probabilistic Soft Logic, the corresponding function
$s(\lambda,d)$ usually has the form $s(\lambda,d)=\lambda\cdot
d^p$ for some real number $p$. Most often, the value $p=1$ is
chosen. Sometimes (but less frequently), the value $p=2$ is
chosen.
\medskip

\noindent{\bf Need to deal with conjunction.} In many rules, the
condition $a_r$ is not a fact, but a conjunction of several facts
$a_{r,1}\,\&\,\ldots\,\&\,a_{r,n_r}$. In other words, such rules
have the form $(a_{r,1}\,\&\,\ldots\,\&\,a_{r,n_r})\to b_r$.

To find the degree $q_r$ to which the rule's condition $a_r$ is
satisfied, it is not sufficient to know the probabilities of all
the facts $a_{r,i}$, we also need to know the dependence between
these random events. For example, if $p(a_{r,1})=p(a_{r_2})=0.5$,
then we can have three different situations:
\begin{itemize}
\item if $a_{r,1}$ and $a_{r_2}$ are independent, then
$p(a_{r,1}\,\&\,a_{r_2})=p(a_{r,1})\cdot p(a_{r_2})=0.25$; \item
if $a_{r,2}$ is equivalent to $a_{r,1}$, then
$p(a_{r,1}\,\&\,a_{r_2})=p(a_{r,1})=0.5$; and \item if $a_{r,2}$
is equivalent to $\neg a_{r,1}$, then
$p(a_{r,1}\,\&\,a_{r_2})=p(a_{r,1}\,\&\,\neg a_{r,1})=0$.
\end{itemize}
In practice, we usually do not know the relation between the
corresponding random events. In this case, we need to come up with
some estimate of the probability
$p(a_{r,1}\,\&\,\ldots\,\&\,a_{r,n_r})$ based only on the known
values $p(a_{r,i})$. Let us denote the function corresponding to
this estimating algorithm by $\wedge(p_1,\ldots,p_n)$. In terms of
this function, once we know the probabilities $p_i\stackrel{\rm
def}{=}p(a_{r,i})$, we estimate the probability of the conjunction
$a_{r,1}\,\&\,\ldots\,\&\,a_{r,n_r}$ as
$\wedge(p_1,\ldots,p_{n_r})$.

Once we know the probabilities $p(a_{r,i})$ of individual events
$a_{r,i}$, the set of possible values of the probability
$p(a_{r,1}\,\&\,\ldots\,\&\,a_{r,n_r})$ is determined by the
Fr\'echet inequalities (see, e.g., \cite{Nelsen1999})
$$\max(p(a_{r,1})+\ldots+p(a_{r,n_r})-(n_r-1),0)\le$$
$$p(a_{r,1}\,\&\,\ldots\,\&\,a_{r,n_r})\le $$
$$\min(p(a_{r,1}),\ldots,p(a_{r,n_r})).\eqno{(2)}$$
\medskip

\noindent{\bf Fr\'echet inequalities explained.} To present a better 
understanding, we now describe where the inequalities (2) come
from.

On the one hand, in every situation in which the conjunction
$a_{r,1}\,\&\,\ldots\,\&\,a_{r,n_r}$ holds, each event $a_{r,i}$
also holds. Thus, for every $i$, the class of all situations in
which the conjunction holds is a subclass of the class of all the
situations in which the $i$-th event $a_{r,i}$ holds. Therefore,
the probability of the conjunction cannot exceed the probability
of the $i$-th event:
$$p(a_{r,1}\,\&\,\ldots\,\&\,a_{r,n_r})\le p(a_{r,i}).$$
The probability of the conjunction is hence smaller than or equal
to $n$ probabilities $p(a_{r,1})$, \ldots, $p(a_{r,n_r})$. Thus,
the probability of the conjunction cannot exceed the smallest of
these $n$ probabilities:
$$p(a_{r,1}\,\&\,\ldots\,\&\,a_{r,n_r})\le
\min(p(a_{r,1}),\ldots,p(a_{r,n_r})).$$ This explains the
right-hand side of the inequality (2).

On the other hand, for every two events $A$ and $B$, we have
$p(A\vee B)=p(A)+p(B)-p(A\,\&\,B)$ and thus, $p(A\vee B)\le
p(A)+p(B)$. By induction, we can conclude that $$p(A_1\vee
\ldots\vee A_{n_r})\le p(A_1)+\ldots+p(A_{n_r})$$ for every
natural number $n_r$. In particular, for the events
$A_i\stackrel{\rm def}{=}\neg a_{r,i}$ for which
$p(A_i)=1-p(a_{r,i})$, we get
$$p((\neg a_{r,1})\vee \ldots\vee (\neg a_{r,n_r}))\le $$
$$(1-p(a_{r,1}))+\ldots+(1-p(a_{r,n_r}))=$$
$$n_r-(p(a_{a,1}+\ldots+p(a_{r,n_r}).$$
Due to de Morgan laws, the disjunction $$(\neg a_{r,1})\vee
\ldots\vee (\neg a_{r,n_r})$$ is equivalent to
$\neg(a_{r,1}\,\&\,\ldots\,\&\,a_{r,n_r}).$ Thus, $$p((\neg
a_{r,1})\vee \ldots\vee (\neg
a_{r,n_r}))=1-p(a_{r,1}\,\&\,\ldots\,\&\,a_{r,n_r}),$$ and the
above inequality takes the form
$$1-p(a_{r,1}\,\&\,\ldots\,\&\,a_{r,n_r})\le $$ $$n_r-(p(a_{a,1})+\ldots+p(a_{r,n_r})).$$
Through simple algebraic manipulation we conclude that
$$p(a_{r,1}\,\&\,\ldots\,\&\,a_{r,n_r})\ge $$ $$p(a_{a,1})+\ldots+p(a_{r,n_r})-(n_r-1).$$
Since the probability is always non-negative, we have
$p(a_{r,1}\,\&\,\ldots\,\&\,a_{r,n_r})\ge 0$. Since the
probability is larger than or equal to the two numbers, it is
therefore larger than the largest of these two numbers:
$$p(a_{r,1}\,\&\,\ldots\,\&\,a_{r,n_r})\ge $$
$$\max(p(a_{a,1})+\ldots+p(a_{r,n_r})-(n_r-1),0).$$ This explains the
first of the inequalities (2).

Thus, Fr\'echet inequalities have been explained.
\medskip

\noindent{\it Comment:} We have derived two inequalities (2). A
natural question is: Can other unrelated inequalities be
similarly derived? It turns out that the two inequalities (2) are
the only limitation on the joint probability
$p(a_{r,1}\,\&\,\ldots\,\&\,a_{r,n_r})$: namely, for every tuple
of values $p_1,\ldots,p_{n_r}$, and for every number $p$ for which
$$\max(p_1+\ldots+p_{n_r}-(n_r-1),0)\le
p\le$$ $$\min(p_1,\ldots,p_{n_r}),$$ we can construct events
$a_{a,1},\ldots,a_{r,n_r}$ with probabilities $p_i=a_{r,i}$ for
which the joint probability is equal exactly to $p$:
$p(a_{r,1}\,\&\,\ldots\,\&\,a_{r,n_r})=p$. \medskip

\medskip{\bf Conclusion - the first desired property of the conjunction operation:}
Fr\'echet inequalities (2) implies that the desired conjunction
operation $\wedge(p_1,\ldots,p_{n})$ should satisfy the inequality
$$\max(p_1+\ldots+p_n-(n-1),0)\le \wedge(p_1,\ldots,p_n)\le $$
$$\min(p_1,\ldots,p_n).\eqno{(3)}$$
\medskip

\noindent{\it Comment:} In many of the Probabilistic Soft Logic formulations 
\cite{Bach2010,Bach2013,Kimmig2012} , usually, the
following conjunction operation is used:
$$\wedge(p_1,\ldots,p_n)=\max(p_1+\ldots+p_n-(n-1),0).\eqno{(4)}$$
This operation clearly satisfies the inequality (3).
\medskip

\noindent{\bf Resulting formulation of the knowledge processing
problems.} Suppose that our knowledge base consists of $R$ rules
of the type $(a_{r,1}\,\&\,\ldots\,\&\,a_{r,n_r})\to b_r$,
$r=1,\ldots,R$, with weights $\lambda_r$.

Then, once we have selected the function $s(\lambda,d)$ and the
conjunction operation $\wedge(p_1,\ldots,p_n)$, we can now find
the probabilities $p(a)$ of different facts $a$ by minimizing the
sum
$$\sum_{r=1}^R
s(\lambda_r,\max(\wedge(p(a_{r,1},\ldots,p(a_{r,n_r})-p(b_r),0)).\eqno{(5)}$$
\medskip

\noindent{\it Comments:} \begin{itemize} \item The need to come up
with some values for $p(a\,\&\,b)$ (and $p(a\vee b)$) when we only
know the probabilities $p(a)$ and $p(b)$ is ubiquitous in
practical applications. In particular, this need underlies the
main ideas behind {\it fuzzy logic}, where the corresponding
conjunction operation $\wedge(p_1,p_2)$ is known as a {\it
t-norm}; see, e.g.,
\cite{Klir1995,Kreinovich1992,Nguyen1997,Nguyen2006,Zadeh1965}.

\item So far, we only described the aspects of the Probabilistic
Soft Logic that enable us to compute the ``most reasonable''
(``most probable'') set of values $p(a)$. In principle, these
techniques also enable us to estimate the probability of other
sets of values $p'(a)\ne p(a)$
\cite{Bach2010,Bach2013,Beltagy2014,Huang2012,Kimmig2012,Memory2012}.
However, as mentioned in \cite{Beltagy2014}, the primary objective
of the Probabilistic Soft Logic is to provide the most reasonable
probabilities. Because of this, in the current paper, we will not
describe how the auxiliary (``second-order'')
probabilities-of-probabilities can be computed.
\end{itemize}
\medskip

\noindent{\bf We need to make sure that the corresponding
computations are efficient.} Our goal is solve practical problems,
and in practice, the number of rules can be large. It is therefore
important to make sure that the corresponding optimization problem
can be solved by a feasible algorithm.

It is known that, in general, optimization problems are NP-hard
(in some formulations, even algorithmically undecidable), but they
become feasible if we restrict ourselves to convex objective
functions; see, e.g., \cite{Vavasis1991}. Moreover, in some
reasonable sense, the class of convex objective functions is the
largest possible class for which optimization is still feasible
\cite{Kearfott2005}.

Thus, we need to make sure that the objective function (5) is
convex. \medskip

\noindent{\bf Conclusion - the second desired property of the
conjunction operation:} Since the function $\max(x,0)$ is convex,
and the superposition of convex functions is always convex, it is
sufficient to make sure:
\begin{itemize} \item that the function $s(\lambda,d)$ is a convex
function of $d\ge 0$, and \item that the conjunction operations
$\wedge(p_1,\ldots,p_n)$ are all convex.
\end{itemize}
\medskip

\noindent{\it Comment:} The choices made in Probabilistic Soft
Logic are indeed convex: \begin{itemize} \item the function
$s(\lambda,d)=\lambda\cdot d^p$ is convex for all $p\ge 1$, and
\item the conjunction operation (4) is convex as well.
\end{itemize}
\medskip

\noindent{\bf What we will do.} We will show that not only is the
conjunction operation (4) convex, \underline{it is the {\it only} possible logical
convex}  \underline{conjunction operation}.
\medskip

\noindent{\it Comment:} The computational problem becomes even
easier when the objective function is not only convex, but also
piece-wise linear, i.e., has the form
$f(x)=\max(\ell_1(x),\ell_2(x),\ldots)$ for several linear
functions $\ell_1(x)$, $\ell_2(x)$, \ldots In this case,
minimizing the objective function is equivalent to solving the
corresponding linear programming problem of minimizing $t$ under
linear constraints $t\ge \ell_1(x)$, $t\ge \ell_2(x)$, \ldots, and
for linear programming problems, efficient algorithms are
available.

The function $\max(a,0)$ is clearly piece-wise linear. Since
superposition of linear functions is linear, superposition of
piece-wise linear functions is piece-wise linear, so to make sure
that the objective function is piece-wise linear, it is sufficient
to make sure:
\begin{itemize} \item that the function
$s(\lambda,d)$ is a piece-wise linear function of $d\ge 0$, and
\item that the conjunction operations $\wedge(p_1,\ldots,p_n)$ are
all piece-wise linear.
\end{itemize}
\medskip

\noindent{\it Comment:} The choices usually made in Probabilistic
Soft Logic are indeed convex and piece-wise linear:
\begin{itemize}
\item the function $s(\lambda,d)=\lambda\cdot d^p$ is piece-wise
linear for $p=1$, and \item the conjunction operation (4) is
piece-wise linear as well.
\end{itemize}

\section{Main Result}

We start with formally defining a conjunction operation.

\noindent{\bf Definition 1.} {\em A function $\wedge:[0,1]^n\to
[0,1]$ is called a {\em logical conjunction operation} if for all possible
inputs $p_1,\ldots,p_n$, it satisfies the inequality
$$\max(p_1+\ldots+p_n-(n-1),0)\le \wedge(p_1,\ldots,p_n)\le $$
$$\min(p_1,\ldots,p_n).\eqno{(3)}$$}
\medskip

\noindent{\bf Proposition 1.} {\em The only convex logical conjunction
operation is the function
$$\wedge(p_1,\ldots,p_n)=\max(p_1+\ldots+p_n-(n-1),0).\eqno{(4)}$$}
\medskip

\noindent{\bf Discussion.} This result explains why the
probabilistic soft logic -- that uses the operation (4) -- allows
for efficient computation of inference, learning, etc.
\cite{Bach2010,Bach2013,Beltagy2014,Huang2012,Kimmig2012,Memory2012}:
indeed, the fact that this operation is convex makes computations
efficient.

This results also shows that the operation (4) is the {\it only}
possible logical convex conjunction operation -- which explains the use of
this operation in Probabilistic Soft Logic.
\medskip

\noindent{\bf Proof.} \medskip

\noindent $1^\circ$. One can easily see that (4) is a logical conjunction
operation, and that it is convex -- as maximum of two linear
(hence convex) functions $p_1+\ldots+p_n-(n-1)$ and~0.
\medskip

\noindent $2^\circ$. Let us now assume that a conjunction
operation $\wedge(p_1,\ldots,p_n)$ is convex. Due to inequalities
(3), for binary vectors $t$, in which each component is 0 or 1, we
have $\wedge(1,\ldots,1)=1$ and $\wedge(t)=0$ for all $t\ne
(1,\ldots,1)$. In particular, for every $i$ from 1 to $n$, we have
$\wedge(t_i)=1$ $t_i=(1,\ldots,1,0,1,\ldots,1)$, where 0 is in the
$i$-th place. \medskip

\noindent $3^\circ$. Let us show that every vector
$p=(p_1,\ldots,p_n)$ with $\sum\limits_{i=1}^n p_i=n-1$ can be
represented as $p=\sum\limits_{i=1}^n a_i\cdot t_i$, with
$a_i=1-p_i$.
\medskip

Indeed, since $p_i\le 1$, we have $a_i=1-p_i\ge 0$ and the
condition $\sum\limits_{i=1}^n p_i=n-1$ implies that
$\sum\limits_{i=1}^n a_i=1$. For each index $j$, the $j$-th
component $s_j$ of the sum $\sum\limits_{i=1}^n (1-p_i)\cdot t_i$
has the form
$$s_j=(1-p_1)+\ldots+(1-p_{j-1})+(1-p_{j+1})+\ldots+(1-p_n)$$
$$=\sum_{i=1}^n
(1-p_i)-(1-p_j).$$ Here, $$\sum_{i=1}^n (1-p_i)=n-\sum_{i=1}^n
p_i=n-(n-1)=1,\eqno{(6)}$$ hence $s_j=1-(1-p_j)=p_j$. The
statement is proven.
\medskip

\noindent $4^\circ$. Let us now prove that every vector $p$ with
$\sum\limits_{i=1}^n p_i\le n-1$ can be represented as a convex
combination of the binary vectors $t$ for which
$\wedge(t)=0$\footnote{As usual, by a convex combination of the
vectors $t_a,\ldots,t_b$, we mean a linear combination in which
all the coefficients are non-negative and add up to 1:
$t=\alpha_a\cdot t_a+\ldots+\alpha_b\cdot t_b$, with $\alpha_a\ge
0$, \ldots, $\alpha_b\ge 0$, and $\alpha_a+\ldots+\alpha_b=1$.}.

Indeed, let us start with such a vector $p=(p_1,\ldots,p_n)$. Let
us consider two cases: $1+\sum\limits_{i=2}^n p_i\ge n-1$ and
$1+\sum\limits_{i=2}^n p_i<n-1$.

In the first case, for the vector $(q_1,p_2,\ldots,p_n)$, where
$q_1=n-1-\sum\limits_{i=2}^n p_i>p_1$, we have
$q_1+p_2+\ldots+p_n=n-1$ and thus, due to Part 3 of this proof,
this vector is a convex combination of the vectors $t$ for which
$\wedge(t)=0$. Here, $0<p_1\le q_1$ thus: \begin{itemize} \item
$p_1$ is a convex combination of numbers 0 and $q_1$. \item So,
the vector $(p_1,\ldots,p_n)$ is a convex combination of the
vectors $(0,p_2,\ldots,p_n)$ and $(q_1,p_2,\ldots,p_n)$. \item The
first vector is a convex combination of binary vectors
$t=(0,\ldots)$ for which $\wedge(t)=0$. \item Thus, the original
vector $p$ is also a convex combination of such vectors.
\end{itemize}

In the second case, the original vector $p$ is a convex
combination of the vectors $(0,p_2,\ldots,p_n)$ and
$(1,p_2,\ldots,p_n)$. Thus, if we prove that the second vector
$(1,p_2,\ldots,p_n)$ can be represented as the desired convex
combination, we will thus prove that the original vector $p$ can
also be represented in this way.

For this new vector $p'=(1,p_2,\ldots,p_n)$, we similarly check
whether raising $p_2$ to 1 will lead to the sum exceeding $n-1$;
if yes, we can similarly complete our proof. If not, then we
reduce the problem to a yet new vector $p''=(1,1,p_3,\ldots,p_n)$,
etc. Eventually, this procedure will stop, since when we already
have $(n-2)$ 1s, then for the corresponding vector
$(1,\ldots,1,p_{n-1},p_n)$ replacement with 1 clearly leads to a
vector $(1,\ldots,1,1,p_n)$ in which the sum of components clearly
exceeds $n-1$. The statement is thus proven.
\medskip

\noindent $5^\circ$. According to Part 4 of this proof, every
vector $p$ with $\sum\limits_{i=1}^n p_i\le n-1$ can be
represented as a convex combination $p=\sum\limits_\alpha
a_\alpha\cdot t_\alpha$ of vectors $t_\alpha$ with
$\wedge(t_\alpha)=0$. Thus, by convexity,
$$\wedge(p)=\wedge\left(\sum\limits_\alpha a_\alpha\cdot
t_\alpha\right)\le \sum\limits_\alpha a_\alpha\cdot
\wedge(t_\alpha)=0.\eqno{(7)}$$ Since the value of the conjunction
operation is always non-negative, this implies that $\wedge(p)=0$
for all such vectors $p$.

So, when $\sum\limits_{i=1}^n p_i\le n-1$, the convex wedge
operation indeed coincides with the desired formula (4). \medskip

\noindent $6^\circ.$ Let us now prove that every vector $p$ with
$\sum\limits_{i=1}^n p_i>n-1$ can be represented as a convex
combination of the vector $t_0\stackrel{\rm def}{=}(1,\ldots,1)$
and vectors $t_i$, namely, that $p=a_0\cdot
t_0+\sum\limits_{i=1}^n a_i\cdot t_i$, with $a_i=1-p_i$ and
$$a_0=1-\sum\limits_{i=1}^n a_i=1-\sum\limits_{i=1}^n
(1-p_i)=\sum_{i=1}^n p_i-(n-1).\eqno{(8)}$$
\medskip

Indeed, since $p_i\le 1$, we have $a_i=1-p_i\ge 0$ and the
condition $\sum\limits_{i=1}^n p_i>n-1$ implies that
$\sum\limits_{i=1}^n a_i=\sum\limits_{i=1}^n (1-p_i)<1$, so
$a_0\ge 0$.

For each index $j$, the $j$-th component $s_j$ of the sum
$a_0+\sum\limits_{i=1}^n a_i\cdot t_i$ has the form
$$s_j=a_0+a_1+\ldots+a_{j-1}+a_{j+1}+\ldots+a_n=\sum_{i=0}^n
a_i-a_j=$$ $$1-a_j=1-(1-p_j)=p_j.\eqno{(9)}$$ The statement is
proven.
\medskip

\noindent $7^\circ$. Due to convexity, when $\sum\limits_{i=1}^n
p_i>n-1$, Part 6 of the proof implies that
$$\wedge(p)=\wedge\left(a_0\cdot t_0+\sum\limits_{i=1}^n
a_i\cdot t_i\right)\le $$ $$a_0\cdot
\wedge(t_0)+\sum\limits_{i=1}^n a_i\cdot \wedge(t_i).\eqno{(10)}$$
Here, $\wedge(t_0)=1$ and $\wedge(t_i)=0$ for all $i$, thus,
$$\wedge(p)\le a_0=\sum_{i=1}^n p_i-(n-1).\eqno{(11)}$$

On the other hand, due to the inequality (3), we have
$$\wedge(p)\ge \max\left(\sum_{i=1}^n p_i-(n-1),0\right).\eqno{(12)}$$
In this case, the difference $\sum\limits_{i=1}^n p_i-(n-1)$ is
positive, so
$$\wedge(p)\ge \sum_{i=1}^n p_i-(n-1).\eqno{(13)}$$
From (11) and (13), we conclude that
$$\wedge(p)=\sum\limits_{i=1}^n p_i-(n-1)=$$ $$
\max\left(\sum_{i=1}^n p_i-(n-1),0\right).\eqno{(14)}$$

So, when $\sum\limits_{i=1}^n p_i>n-1$, the convex wedge operation
also coincides with the desired formula (4). The proposition is
proven.

\section{What If We Take into Account that Human ``And'' Is
Somewhat Different from the Formal Conjunction}

\noindent{\bf Human ``and'' is somewhat different from formal
``and''.} In formal logic, if one of the conditions
$a_1,\ldots,a_n$ is not satisfied, then the whole conjunction
$a_1\,\&\,\ldots\,\&\,a_n$ is false. For example, when $a_1$ and
$a_2$ are true, but $a_3$ is false, we get $\wedge(1,0,1)=0$.

In human reasoning, this is not always so. For example, when a
department tries to hire a new faculty member, the usual
requirement is that this person should be a good researcher
($a_1$), a very good teacher ($a_2$), and a good colleague ($a_3$.
Ideally, this is who we want to hire: a person who satisfies the
property $a_1\,\&\,a_2\,\&\,a_3$.

Let us now assume that one of the candidates is an excellent
researcher who is going to be a very good colleague, but whose
teaching skills are not so good, i.e., for whom $a_1=1$, $a_2=0$,
and $a_3=0$. Formally, in this case, one of the conditions is not
satisfied, so the conjunction $a_1\,\&\,a_2\,\&\,a_3$ is false.
However, in reality, this person has a reasonable chance of being
hired -- meaning that, according to our human reasoning, the
original ``and''-rule is to some extent satisfied, i.e.,
$\wedge(1,0,1)>0$.
\medskip

\noindent{\bf How to formally describe this difference.} To
describe this difference, researchers have proposed several
conjunction operations for which $\wedge(1,0,1)> 0$; see, e.g.,
\cite{Kreinovich2004,Trejo2002,Zimmermann1980}.

In the context of Probabilistic Soft Logic, the paper
\cite{Beltagy2014} uses an operation
$$\wedge(p_1,\ldots,p_n)=\displaystyle\frac{1}{n}\cdot
\sum\limits_{i=1}^n p_i\eqno{(15)}$$ for which also
$\wedge(1,0,1)=2/3>0$.
\medskip

\noindent{\bf Which conjunction operation should we use: analysis
of the problem.} As we have mentioned earlier, the
easiest-to-process convex objective functions are piece-wise
linear functions, i.e., maxima of several linear functions. The
fewer linear functions we have, the easier it is to process this
problem. For the operation (4), we have two linear functions, so
let us use two linear functions here as well.

Also, the simpler one of these functions, the easier it is to
solve the corresponding linear programming problem. In the
formal-``and'' case, one of this functions was 0, so let us use 0
here as well. Thus, we are looking for a conjunction operation of
the type $\wedge(p)=\max(\ell(p),0)$ for some linear function
$\ell(p)=c_0+\sum\limits_{i=1}^n c_i\cdot p_i$.

In general, ``$a_1$ and $a_2$'' means the same as ``$a_2$ and
$a_1$''. Thus, the ``truth value'' $\wedge(p_1,\ldots,p_n)$ should
not change if we simply permute the inputs. Hence, the
corresponding linear function should be permutation-invariant,
which implies that $c_1=c_2=\ldots=c_n$.

We should have $\wedge(0,\ldots,0)=0$ and $\wedge(1,\ldots,1)=1$.
The fact that $\wedge(1,\ldots,1)>\wedge(0,\ldots,0)$ implies that
$c_1>0$. The first condition $\wedge(0,\ldots,0)=0$ then means
that $c_0\le 0$, and the second condition means that $c_0+n\cdot
c_1=1$. So, $c_0=1-n\cdot c_1$, and the requirement that $c_0\le
0$ means that $c_1\ge \displaystyle\frac{1}{n}$. Thus, we arrive
at the following recommendation.
\medskip

\noindent{\bf What we propose.} We propose to use the following
conjunction operation
$$\wedge(p_1,\ldots,p_n)=$$ $$\max\left(c_1\cdot \left(\sum_{i=1}^n
p_i\right)-(n\cdot c_1-1),0\right),\eqno{(16)}$$ where
$c_1\in\left[\displaystyle\frac{1}{n},1\right].$
\medskip

\noindent{\bf Properties of the proposed operation.} The proposed
operation is {\it convex} -- and thus, similarly to the operation
(4), it leads to efficient optimization hence to efficient
inference, learning, etc.

However, this operation is {\it no longer associative}. For
example, for $c_1=\displaystyle\frac{1}{n}$, the operation (16)
turns into the arithmetic average (15). For the arithmetic
average,
$$\wedge(0,\wedge(0.5,1))=\wedge\left(0,\frac{0.5+1}{2}\right)=
\wedge(0,0.75)=$$ $$\frac{0+0.75}{2}=0.375,$$ while
$$\wedge(\wedge(0,0.5),1)=\wedge\left(\frac{0+0.5}{2},1\right)=
\wedge(0.25,1)=$$ $$\frac{0.25+1}{2}=0.625\ne 0.375.$$
\medskip

\noindent{\bf The proposed family of operations contains both
currently used operations of Probabilistic Soft Logic.} The
parameter $c_1$ can take all possible values from
$\displaystyle\frac{1}{n}$ to 1.
\begin{itemize} \item On one extreme, for $c_1=1$, we get the usual
formal-``and''-based operation~(4) used in 
\cite{Bach2010,Bach2013,Kimmig2012}. 

\item At the other extreme,
when $c_1=\displaystyle\frac{1}{n}$, we get the arithmetic average
operation (15) used in \cite{Beltagy2014}. \end{itemize}
\medskip

\noindent{\bf Potential advantage of the proposed family of
operations.} In general, the family (16) provides an additional
parameter $c_1$ that we can adjust to hopefully get an even better
match with human reasoning -- while still retaining convexity and
thus, retaining computational efficiency.

\section{Conclusion}
Probabilistic Soft Logic (PSL) is a probabilistic formalism that can
be used in learning, representing and reasoning with uncertain 
and possibly inconsistent (when weights are not taken into account) 
knowledge, and it seems to scale up better than its alternatives.
A key aspect of PSL is its use of continuous truth values. This necessitates 
alternatives to boolean conjunction operations and at least two such operations
have been mentioned in the literature.

In this paper we have presented formal justifications 
of two different conjunction operations used in 
various PSL formulations and applications.  We first showed (in Proposition 1) that 
the conjunction operations used in the original PSL \cite{Bach2010,Bach2013,Kimmig2012}
is unique in that it is the only conjunction operation that satisfies a
set of desired properties. Next we presented a family of conjunction operations (16)
that satisfy a smaller set of desired properties. We show that this family of operations
contain both the operation used in the original PSL \cite{Bach2010,Bach2013,Kimmig2012} and the
operation used in \cite{Beltagy2014}. Our presentation of a family of operations
gives us additional conjunction operations which may come in handy in other situations.

\newpage

\bibliographystyle{aaai}
\bibliography{vladik}

\end{document}